\documentclass[prl, twocolumn]{revtex4}

\usepackage{amsfonts, amssymb, amsmath}
\usepackage{bm, bbm}

\newcommand{\sla}[1]{\setbox0=\hbox{$#1$}       
   \dimen0=\wd0                                 
   \setbox1=\hbox{/} \dimen1=\wd1               
   \ifdim\dimen0>\dimen1                        
      \rlap{\hbox to \dimen0{\hfil/\hfil}}      
      #1                                        
   \else                                        
      \rlap{\hbox to \dimen1{\hfil$#1$\hfil}}   
      /                                         
   \fi}                                         %

\newcommand{\eql}[1]{\label{eq:#1}}
\newcommand{\eq}[1]{(\ref{eq:#1})}

\newcommand{\Ref}[1]{Ref.$\,$\cite{#1}}

\newcommand{\del}{\partial}

\newcommand{\wba}[1]{\overline{#1}}

\newcommand{\fdot}{\!\cdot\!}

\newcommand{\cB}{\mathcal{B}}
\newcommand{\cC}{\mathcal{C}}
\newcommand{\cL}{\mathcal{L}}
\newcommand{\cO}{\mathcal{O}}

\newcommand{\Z}{\mathbb{Z}}
\newcommand{\U}{\mathrm{U}}

\newcommand{\cc}{\text{c.c.}}

\newcommand{\de}{\delta}
\newcommand{\ga}{\gamma}

\newcommand{\sg}{\sigma}

\newcommand{\D}{\mathrm{D}}
\newcommand{\M}{\mathrm{M}}
\newcommand{\T}{\mathrm{T}}

\newcommand{\dps}[1]{\displaystyle{#1}}

\begin{document}

\title{Effective Field Theory for a Heavy Majorana Fermion}

\author{Karoline K\"opp}
\affiliation{Department of Physics, Florida State University, Tallahassee, FL 32306, USA} 
\author{Takemichi Okui}
\affiliation{Department of Physics, Florida State University, Tallahassee, FL 32306, USA} 

\begin{abstract}
We formulate an effective theory for systems containing a heavy Majorana fermion, 
such as bound states of a long-lived gluino.  This ``Majorana HQET'' has the same 
degrees of freedom as the well-studied Dirac HQET\@.  It respects an emergent $\U(1)$ 
symmetry despite the fundamental absence of a $\U(1)$ for Majorana fermions.  
Reparameterization invariance works identically in the two HQETs.  Remarkably, 
while a Dirac HQET may or may not be charge conjugation symmetric, a charge 
conjugation symmetry emerges in all Majorana HQETs, potentially offering low 
energy probes to distinguish the two theories.
\end{abstract}

\maketitle

While all fermions in the standard model (SM) --- except possibly for neutrinos --- 
are Dirac fermions, massive Majorana fermions are quite common in models beyond the SM\@.  
A particularly well-known example is the gluino, a heavy color-octet Majorana fermion
in minimal supersymmetric extensions of the SM~\cite{Martin:1997ns}. 
If the gluino's lifetime is much longer than the timescale of QCD confinement  
$\dps{\sim 10^{-24}\>\text{s}}$ (as happens in split 
supersymmetry~\cite{ArkaniHamed:2004fb, Giudice:2004tc}, for instance), 
it will form a heavy QCD bound state with gluons and quarks~\cite{Farrar:1978xj}. 
Such bound states, called R-hadrons, are actively searched for by experiments at 
the Large Hadron Collider~\cite{Aad:2011hz, Khachatryan:2011ts}.  Therefore, it is 
important to develop a theoretical framework to systematically analyze the system 
of a heavy Majorana fermion interacting with light particles.
   
Heavy quark effective theory 
(HQET)~\cite{Eichten:1989zv, Grinstein:1990mj, Georgi:1990um, Falk:1990yz} 
is an effective field theory originally developed for studying the properties 
of mesons containing a heavy quark such as the $b$ quark.  
The standard formulation of HQET, however, presumes that the fermion is Dirac, 
and naively applying it to Majorana fermions has led \Ref{Chakraverty:1995tb} 
to propose that the celebrated 
spin symmetry~\cite{Shifman:1986sm, Shifman:1987rj, Isgur:1989vq, Isgur:1989ed} 
of HQET is lost in the Majorana case (which fortunately is false, as we will show).

In this paper, we will derive the HQET for a Majorana fermion and study its basic properties.  
First, we will show that Majorana HQET has the same effective degrees of freedom
and propagators as the well-known Dirac HQET\@.
This agrees with the intuition that, at scales below the threshold for particle-antiparticle 
pair creation, 
we should not be able to tell whether or not the antiparticle is distinct from the particle. 

We will further sharpen the similarities between Majorana and Dirac HQETs by
demonstrating that Majorana HQET possesses an emergent $\U(1)$ symmetry 
as if it came from a Dirac fermion, even though Majorana fermions by definition cannot 
have any fundamental $\U(1)$ symmetry.
The emergence of the effective $\U(1)$ should in fact be viewed as a consistency check, 
since Majorana HQET trivially conserves particle number as being an effective theory 
for one-fermion states. 
We also show that reparameterization 
invariance~\cite{Luke:1992cs, Chen:1993sx, Kilian:1994mg, Sundrum:1997ut} works in the same way
in Majorana and Dirac HQETs.

Despite these common features, the two effective theories {\em are} different.  
We will show that any Majorana HQET must be also equipped with an emergent charge conjugation 
symmetry, even if the full theory lacks a charge conjugation symmetry.  
This effective charge conjugation symmetry is an exact, intrinsic property of Majorana HQET
that reflects the absence of particle-antiparticle distinction in the original Majorana fermion. 
This contrasts to the Dirac case, where a Dirac HQET
would have a charge conjugation symmetry only if the full theory has one.
Therefore, this potentially provides low energy probes to distinguish
between R-hadrons containing a Majorana gluino and those 
containing a Dirac gluino that appears in many non-minimal models of 
supersymmetry, for example.

\section{Degrees of Freedom}
We first derive the kinetic and mass terms of Majorana HQET\@.
We will see that the quadratic terms of the Majorana HQET lagrangian exactly agree
with those of Dirac HQET, so the degrees of freedom of the two effective theories 
are identical in content and propagate in the same manner.  

To perform a blow-by-blow comparison between the Majorana and Dirac cases,
let's consider a Majorana fermion $\psi_\M(x)$ and a Dirac fermion $\psi_\D(x)$ with 
the same mass $m$ and
identical quantum numbers under all symmetries, except for $\U(1)_\D$, the very $\U(1)$
that defines the Dirac fermion by providing particle number conservation. 
Being Majorana, $\psi_\M(x)$ obeys the constraint%
\begin{equation}
\psi_\M(x) = \cB \psi_\M^\ast(x) 
\eql{majo},
\end{equation}
with the Majorana conjugation matrix $\cB$ satisfying%
\begin{equation}
\cB {\ga^\mu}^\ast \cB^{-1}
= -\ga^\mu
\,,\quad
\cB^\ast = \cB^{-1}
\,,\quad
\cB^\T = \cB
\,.
\end{equation}

In HQET, we are interested in the states of a single $\psi$ particle and arbitrary
numbers of other particles with masses $\dps{\ll m}$, where interactions are 
transferring only small momenta ($\dps{\ll m}$) to the $\psi$ particle, hence 
never exciting another $\psi$ particle.
Thus, perturbation theory should begin with 
a free $\psi$ particle with a 4-momentum $mv$, where $v$ 
is a timelike 4-velocity 
with $\dps{v^0>0}$ and $\dps{v\fdot v = 1}$.  
This state is described by the following solutions of the free Dirac equations 
$\dps{(i\sla{\del}- m)\psi_{\D,\M} = 0}$:%
\begin{eqnarray}
&&
\psi_\D(x)
= e^{- imv\cdot x} u_v
\eql{dirac_soln}\,,\\
&&
\psi_\M(x) 
= e^{-imv\cdot x} u_{v} 
+ e^{imv\cdot x} \cB u_{v}^\ast
\eql{majo_soln}\,,
\end{eqnarray}
where $u_v$ is a constant spinor obeying the constraint
$\dps{\sla{v} u_v = u_v}$.

Now, turning on the interactions, we would like to parameterize
the fluctuations of $\psi_{\D,\M}(x)$ around the solutions~\eq{dirac_soln} and~\eq{majo_soln}~%
\footnote{In interacting theories, the mass parameter $m$ 
depends on the renormalization scheme.  The standard, convenient (but not mandatory) choice
is to set the $h_v$ mass term to zero in the HQET lagrangian
to be discussed below.
}.
Away from these free-particle solutions, the above constant spinor $u_v$ with only the $\sla{v}=1$ 
component has to be replaced by an arbitrary $x$-dependent spinor with 
both $\dps{\sla{v} = \pm1}$ components.
Thus, we are led to the following changes of variables:%
\begin{eqnarray}
&&
\psi_\D(x)
= e^{-imv\cdot x} [ h_v(x) + H_v(x) ] 
\eql{dirac:h_and_H}\,,\\
&&
\begin{aligned}
\psi_\M(x) 
= {}&
e^{-imv\cdot x} [ h_v(x) + H_v(x) ] \\ 
&
+ e^{imv\cdot x} \cB [ h_v^\ast(x) + H_v^\ast(x) ]
\,.
\end{aligned}  
\eql{majo:h_and_H}
\end{eqnarray}
where $h_v(x)$ and $H_v(x)$ parameterize the $\dps{\sla{v}=+1}$ and $\dps{\sla{v}=-1}$ 
components, respectively, i.e.:%
\begin{equation}
\sla{v} h_v(x) = h_v(x)
\,,\quad     
\sla{v} H_v(x) = -H_v(x)
\,.
\end{equation}
Note that the Majorana condition~\eq{majo} is kept manifest in the 
parameterization~\eq{majo:h_and_H}.  

In terms of these new variables, the quadratic part of the full lagrangian 
for the Dirac theory can be rewritten as%
\begin{equation}
\begin{split}
\hspace{-0.8ex}
\cL^{(2)}_\text{D,full}
= {}& 
\wba{\psi}_\D (i\sla{\del} - m) \psi_\D
\\
= {}&
i\bar{h}_v \, v\fdot\del h_v
-i\overline{H}_v \, v\fdot\del H_v 
-2m \, \overline{H}_v H_v 
\\
&
+( i\bar{h}_v \,\sg^{\mu\nu} v_\mu \del_{\perp\nu} H_v 
+ \cc )
\,,
\end{split}
\eql{dirac_full} 
\end{equation}
where $\dps{\sg^{\mu\nu} \equiv [\ga^\mu, \ga^\nu]/2}$ and
$\dps{\del_{\perp\mu} \equiv \del_\mu - (v\fdot\del)v_\mu}$.
(Needless to say, we have suppressed the kinetic and mass terms of other fields, e.g.,
the gluon.)
Similarly, the quadratic part of the full Majorana lagrangian can be rewritten as
\begin{equation}
\begin{split}
\cL^{(2)}_\text{M,full}
= {}&
\frac12\, \wba{\psi}_\M (i\sla{\del} - m) \psi_\M
\\
= {}&
\cL^{(2)}_\text{D,full}
+\bigl( e^{-2imv\cdot x} P(h_v, H_v) + \cc \bigr)
\,,
\end{split}
\eql{majo_full} 
\end{equation}
where $P(h_v, H_v)$ is a quadratic polynomial of $h_v(x)$, $H_v(x)$
and their derivatives, 
with constant coefficients (i.e., no explicit $x$-dependence like $e^{-2imv\cdot x}$ 
inside $P$), which stems from picking up the $e^{-imv\cdot x}$ term of~\eq{majo:h_and_H} twice.  

We are now ready to construct effective theories that reproduce these full theories when we
specialize in the states of a single $\psi$ fermion with momentum near $mv$, 
plus arbitrary numbers of other light particles with masses and momenta $\dps{\ll m}$. 
The effective theories can be derived from the full theories by restricting the fields 
$h_v(x)$ and $H_v(x)$ to only contain ``slowly varying'' modes
with wavelengths $\dps{\gg m^{-1}}$.
This does not correctly take into account loop diagrams, where the 
loop momenta go to infinity, but these ``mistakes'' can be fully corrected in the effective theory 
by adding local operators with appropriate coefficients (i.e.~``matching'').

For the purpose of understanding the field content and the form of propagators in 
the effective theory, 
tree-level matching is sufficient. Thus, the quadratic terms of the 
effective lagrangian for the Dirac case can simply be given by the lagrangian~\eq{dirac_full} 
with the understanding that $h_v$ and $H_v$ only contain modes with
wavelengths $\dps{\gg m^{-1}}$.  So,%
\begin{equation}
\begin{split}
\cL^{(2)}_\text{D,eff}
= {}&
i\bar{h}_v \, v\fdot\del h_v
-i\overline{H}_v \, v\fdot\del H_v 
-2m \, \overline{H}_v H_v 
\\
&
+( i\bar{h}_v \,\sg^{\mu\nu} v_\mu \del_{\perp\nu} H_v 
+ \cc )
\,.
\end{split}
\eql{dirac_eff}
\end{equation}
For the Majorana case, the effective lagrangian obtained this way 
from the lagrangian~\eq{majo_full} appears to contain the additional 
terms $\dps{e^{-2imv\cdot x}P(h_v, H_v)+\cc}$.  
These terms actually vanish once integrated over spacetime to obtain the
effective action.
In momentum space, they are transformed into $\delta$-functions of 
the form $\delta^4(2mv+k)$, where $k$
is the momentum carried by $P(h_v, H_v)$. However, since $h_v(x)$ and $H_v(x)$ 
in the effective theory are restricted to be slowly varying, 
$k$ is necessarily $\ll mv$ and the $\delta$-functions simply vanish.
Therefore, we obtain%
\begin{equation}
\begin{split}
\cL^{(2)}_\text{M,eff}
= \cL^{(2)}_\text{D,eff}
\,.
\end{split}
\eql{majo_eff}
\end{equation}
We conclude that the field content and propagators of Majorana HQET are identical to
those of Dirac HQET\@.  
This in particular implies that the spin symmetry of HQET in the limit of decoupling $H_v$ 
is intact in Majorana HQET, contrary to the claim made in \Ref{Chakraverty:1995tb}.

\section{Symmetries}
Next, we would like to compare symmetries of the two effective theories.
First, the full Dirac and Majorana theories we started with have identical symmetries by assumption, 
except for $\U(1)_\D$ of the Dirac fermion.
Clearly, all these symmetries of the full theories are passed down to their respective HQETs.

In addition, it is well-known that HQET possesses an emergent ``gauge symmetry'' called 
reparameterization invariance 
(RPI)~\cite{Luke:1992cs, Chen:1993sx, Kilian:1994mg, Sundrum:1997ut}, 
which is a redundancy in the HQET description 
that choosing a different $v$ should not change the physics.  
Below, we show that RPI works in exactly the same way 
in the Majorana and Dirac HQETs.

Furthermore, in Majorana HQET, the $\U(1)_\D$ global symmetry emerges as an exact symmetry.
Intuitively, this is because, being an effective theory of one-particle states, 
Majorana HQET trivially conserves particle number, even though the full Majorana theory 
has no particle number conservation.  Below, we will demonstrate explicitly 
how this $\U(1)$ arises in Majorana HQET\@.   

Remarkably, we will see yet another symmetry emerging in Majorana HQET, so the symmetry content 
of Majorana HQET is actually larger than that of Dirac HQET\@.
This symmetry is an effective charge conjugation symmetry that reflects the very Majorana nature 
of the full theory, namely the absence of particle/antiparticle distinction.  Thus, this is 
an exact symmetry of any Majorana HQET, even without a charge conjugation symmetry in the full
theory.  In contrast, charge conjugation symmetry is optional
for Dirac fermions and for Dirac HQETs.
This emergent charge conjugation symmetry forbids a class of operators in Majorana HQET that are 
allowed in Dirac HQET\@.  Therefore, discovering the effects of those operators in experiment 
can tell us that the underlying fermion must be Dirac.

\subsection{Reparameterization Invariance}

The RPI redundancy in choosing $v$ is manifest in the relations~\eq{dirac:h_and_H} 
and~\eq{majo:h_and_H}, where the left-hand sides simply have no reference to $v$.
For the Dirac case, this readily implies that the fields labelled by $v' \equiv v+\de v$ must be
related to those labelled by $v$ as%
\begin{equation}
e^{- im v' \cdot x} ( h_{v'} + H_{v'} )
= e^{- im v \cdot x} ( h_v + H_v )
\eql{Dirac_RPI}\,,
\end{equation}
provided that $m\de v \ll m$ so that we maintain the requirement
that $h_{v'}(x)$ and $H_{v'}(x)$ should vary slowly in $x$ with wavelengths
$\dps{\gg m^{-1}}$, just like $h_v(x)$ and $H_v(x)$ themselves.
This is exactly the form of RPI proven to be valid to all orders in Dirac HQET by
\Ref{Sundrum:1997ut}.  

For the Majorana case, the $v$-independence of the left-hand side of~\eq{majo:h_and_H} 
implies that%
\begin{equation}
\begin{split}
{}&
e^{-imv'\cdot x} ( h_{v'} + H_{v'} )
+ e^{imv'\cdot x} \cB ( h_{v'}^\ast + H_{v'}^\ast )
\\ 
= {}&
e^{-imv\cdot x} ( h_v + H_v )
+ e^{imv\cdot x} \cB ( h_v^\ast + H_v^\ast )
\,.
\end{split}
\eql{majo_RPI}
\end{equation}
This naively appears different from the Dirac RPI~\eq{Dirac_RPI}. 
However, since the fields $h_{v'}$, $H_{v'}$, $h_v$ and $H_v$ 
only carry momenta $\dps{\ll mv}$, the 1st term on each side of~\eq{majo_RPI} only contains
positive frequency modes, while the 2nd term on each side only contains 
negative frequency modes.  Thus, the 1st and 2nd terms are linearly independent and can
be equated separately.  This then gives a relation identical to the Dirac RPI~\eq{Dirac_RPI}.
We conclude that RPI works identically in Dirac and Majorana HQETs.

\subsection{Emergent $\U(1)_\D$ in Majorana HQET}

The equality~\eq{majo_eff} trivially implies that the quadratic part $\cL^{(2)}_\text{M,eff}$ 
respects the same $\U(1)_\D$ global symmetry as the Dirac counterpart, 
with both $h_v$ and $H_v$ carrying a unit charge, 
even though the full Majorana theory~\eq{majo_full} possesses no $\U(1)$ symmetry.
We will now show that $\U(1)_\D$ is an exact symmetry
of the entire Majorana HQET lagrangian, not only of the quadratic part,
as expected from the fact that Majorana HQET is a theory of a fixed number of 
$\psi$ particles (namely, one). 

The emergence of $\U(1)_\D$ can be demonstrated elegantly by using RPI\@.
Since we have shown that Majorana RPI is the same as Dirac RPI, Majorana HQET 
has the same ``RPI invariant'' as the Dirac RPI\@. Namely, the RPI relation~\eq{Dirac_RPI} 
implies that the linear combination%
\begin{equation}
\chi(x) 
\equiv e^{- imv\cdot x} [h_v(x) + H_v(x)] 
\eql{RPIinvariants}
\end{equation}
(not to be conceptually confused with the full-theory Dirac field $\psi_\D$) 
are invariant under $v \to v+\de v$.  
Therefore, RPI can be made manifest by writing the lagrangian solely in terms of $\chi$.

Now, since HQET is an effective theory for single-$\psi$-particle states, we only need to look at
operators that are bilinear in $\chi$.  
Thus, all operators in Majorana HQET are in either one of the following forms:%
\begin{equation}
\wba{\chi} \cO \chi
\,,\quad 
\chi^\T \cC \cO \chi
\eql{emergentU1}\,,
\end{equation}
where $\cO$ contains $\ga$ matrices, derivatives and other light fields in the theory,
while the charge conjugation matrix $\cC$ satisfies%
\begin{equation}
\cC {\ga^\mu}^\T \cC^{-1}
= -\ga^\mu
\,,\quad
\cC^\ast = -\cC^{-1}
\,,\quad
\cC^\T = -\cC
\,.
\end{equation}
The operators of the 1st type in~\eq{emergentU1} 
are invariant under $\U(1)_\D$, while those of the 2nd type are not.
Notice, however, that the latter do not exist in the effective action, 
because they contain rapid oscillations $e^{\pm 2imv\cdot x}$ and thus
vanish under the integration over spacetime, just like what happened to $P(h_v, H_v)$ previously.
Therefore, $\U(1)_\D$ is indeed respected by all operators in Majorana HQET\@.

\subsection{Emergent $\Z_2$ Symmetry in Majorana HQET}
Notice that the right-hand side of~\eq{majo:h_and_H}
is unchanged by the simultaneous operations of%
\begin{equation}
v \leftrightarrow -v
\,,\quad
h_v \leftrightarrow \cB h_v^\ast
\,,\quad
H_v \leftrightarrow \cB H_v^\ast
\,.
\eql{emergentZ2}
\end{equation}
Majorana HQET must be invariant under this charge conjugation operation, because, like RPI,
this is a redundant operation, doing nothing to the full-theory variable $\psi_\M$.
This redundancy makes sense intuitively, because the original Majorana fermion does 
not distinguish particle and antiparticle so it should not ``care'' whether we have chosen $v$ 
or $-v$ to write the effective theory.

In contrast, Dirac HQET does not in general respect the charge conjugation symmetry~\eq{emergentZ2},  
unless the full theory happens to be invariant under the charge conjugation 
$\psi_\D \leftrightarrow \cB \psi_\D^\ast$.
It should be stressed, however, that such charge conjugation symmetry may or may not be there 
in a given Dirac theory, while any Majorana HQET must have the symmetry~\eq{emergentZ2}, 
regardless of the presence or absence of charge conjugation symmetry in the full theory,
as it is merely a redundancy of the formulation, similarly to RPI\@.

The emergent charge conjugation symmetry~\eq{emergentZ2}
imposes nontrivial constraints on the structure of Majorana HQET lagrangian.  
As a theoretical illustration, consider a toy model consisting of a heavy, color-octet 
Majorana fermion $\psi_\M = \psi_\M^a T^a$ ($a=1$, $\cdots$, $8$), where 
$T^a \equiv \lambda^a/2$ with the Gell-Mann matrices $\lambda^a$.
Suppose that there is also a color-octet real scalar $\phi = \phi^a T^a$ with mass much
lighter than $\psi$. 
Then, in the full theory, symmetry permits the Yukawa interaction of the form%
\begin{equation}
d^{abc} \, \phi^a (\psi_\M^b)^\T \cC \psi_\M^c
\,,
\end{equation}
with the totally symmetric $\dps{d^{abc} \propto \mathrm{tr}[T^a \{T^b,T^c\}]}$.  
The analogous term 
$\dps{f^{abc} \, \phi^a (\psi_\M^b)^\T \cC \psi_\M^c}$ with the totally antisymmetric
$\dps{f^{abc} \propto \mathrm{tr}[T^a[T^b,T^c]]}$ vanishes simply due to
the {\em algebraic} identity reflecting the Fermi-Dirac statistics, 
$\dps{(\psi_\M^b)^\T \cC \psi_\M^c = (\psi_\M^c)^\T \cC \psi_\M^b}$, 
without owing to any symmetry.
This interaction matches at tree level onto the HQET operator%
\begin{equation}
d^{abc} \, \phi^a \bar{h}_v^b h_v^c + \cdots
\,,
\end{equation}
where the ellipses indicate similar terms containing $H_v$.

In stark contrast, if the fermion were a color-octet Dirac fermion $\psi_\D = \psi_\D^a T^a$, 
both types of the interactions would be allowed:%
\begin{equation}
d^{abc} \, \phi^a \wba{\psi}_\D^b \psi_\D^c
+ f^{abc} \, \phi^a \wba{\psi}_\D^b \psi_\D^c
\,,
\end{equation}
since this time $\wba{\psi}_\D^b \psi_\D^c \neq \wba{\psi}_\D^c \psi_\D^b$.
These interactions match at tree level onto the HQET operators%
\begin{equation}
d^{abc} \, \phi^a \bar{h}_v^b h_v^c 
+ f^{abc} \, \phi^a \bar{h}_v^b h_v^c
+ \cdots
\,,\eql{dabc+fabc}
\end{equation}
where the ellipses indicate similar terms containing $H_v$.

Now, without the emergent charge conjugation symmetry~\eq{emergentZ2}, 
one would think that the $f^{ijk}$ operator like in~\eq{dabc+fabc} should be also generated
in Majorana HQET at loop level, 
since Majorana HQET would have exactly the same symmetries and same degrees of freedom 
as Dirac HQET~%
\footnote{If we assume renormalizability in the full Majorana theory, 
we would have the accidental charge conjugation symmetry
$\dps{\psi^a T^a \to \pm\psi^a (-T^a)^\T}$, $\dps{\phi^a T^a \to -\phi^a (-T^a)^\T}$ 
and $\dps{G^a_\mu T^a \to G^a_\mu (-T^a)^\T}$, with 
$G_\mu$ being the gluon field.  
Then, correspondingly in Majorana HQET, the symmetry  
$\dps{h_v^a T^a \to \pm h_v^a (-T^a)^\T}$, etc., would forbid the $f^{abc}$ term.
However, this accidental symmetry can easily be broken by non-renormalizable interactions, e.g., 
$\dps{f^{abc} d^{cde} \, \phi^a (\del_\mu \phi^b) \wba{\psi}^d \ga^\mu \psi^e}$,
which we assume to be present.}. 
However, under the symmetry~\eq{emergentZ2}, 
the $d^{abc}$ operator is allowed but the $f^{abc}$ operator is not, because 
the operation~\eq{emergentZ2} gives%
\begin{equation}
\bar{h}_v^b h_v^c
\longrightarrow
(h_v^b)^\T \cB^{-1}\ga^0 \cB h_v^{c\ast} 
= 
\bar{h}_v^c h_v^b
\,,
\end{equation}
so the $f^{abc}$ operator would change the sign.
Therefore, the emergent charge conjugation of Majorana HQET guarantees that the $f^{ijk}$ operator 
is never generated at any order in loop expansion.  

We conclude that emergent charge conjugation symmetry of Majorana HQET offers 
the interesting opportunity to tell apart Dirac and Majorana
fermions from only low energy measurements performed 
at scales much below the fermion mass threshold.
Namely, probing the presence of the HQET operators forbidden by~\eq{emergentZ2} 
can rule out the possibility that the fermion is Majorana.          

\vskip.1in
\noindent
{\em Acknowledgment:} We thank M.~Luty, A.~Nelson and R.~Sundrum for comments on the draft.
We also thank the anonymous referee for pointing out an error in the first version 
where the RPI literature was cited with the wrong reference numbers.
This work is supported by DOE grant No.~DE-FG02-97ER41022.




\begin{thebibliography}{19}
\expandafter\ifx\csname natexlab\endcsname\relax\def\natexlab#1{#1}\fi
\expandafter\ifx\csname bibnamefont\endcsname\relax
  \def\bibnamefont#1{#1}\fi
\expandafter\ifx\csname bibfnamefont\endcsname\relax
  \def\bibfnamefont#1{#1}\fi
\expandafter\ifx\csname citenamefont\endcsname\relax
  \def\citenamefont#1{#1}\fi
\expandafter\ifx\csname url\endcsname\relax
  \def\url#1{\texttt{#1}}\fi
\expandafter\ifx\csname urlprefix\endcsname\relax\def\urlprefix{URL }\fi
\providecommand{\bibinfo}[2]{#2}
\providecommand{\eprint}[2][]{\url{#2}}

\bibitem[{\citenamefont{Martin}(1997)}]{Martin:1997ns}
\bibinfo{author}{\bibfnamefont{S.~P.} \bibnamefont{Martin}}
  (\bibinfo{year}{1997}), \eprint{hep-ph/9709356}.

\bibitem[{\citenamefont{Arkani-Hamed and
  Dimopoulos}(2005)}]{ArkaniHamed:2004fb}
\bibinfo{author}{\bibfnamefont{N.}~\bibnamefont{Arkani-Hamed}}
  \bibnamefont{and}
  \bibinfo{author}{\bibfnamefont{S.}~\bibnamefont{Dimopoulos}},
  \bibinfo{journal}{JHEP} \textbf{\bibinfo{volume}{0506}}, \bibinfo{pages}{073}
  (\bibinfo{year}{2005}), \eprint{hep-th/0405159}.

\bibitem[{\citenamefont{Giudice and Romanino}(2004)}]{Giudice:2004tc}
\bibinfo{author}{\bibfnamefont{G.}~\bibnamefont{Giudice}} \bibnamefont{and}
  \bibinfo{author}{\bibfnamefont{A.}~\bibnamefont{Romanino}},
  \bibinfo{journal}{Nucl.Phys.} \textbf{\bibinfo{volume}{B699}},
  \bibinfo{pages}{65} (\bibinfo{year}{2004}), \eprint{hep-ph/0406088}.

\bibitem[{\citenamefont{Farrar and Fayet}(1978)}]{Farrar:1978xj}
\bibinfo{author}{\bibfnamefont{G.~R.} \bibnamefont{Farrar}} \bibnamefont{and}
  \bibinfo{author}{\bibfnamefont{P.}~\bibnamefont{Fayet}},
  \bibinfo{journal}{Phys.Lett.} \textbf{\bibinfo{volume}{B76}},
  \bibinfo{pages}{575} (\bibinfo{year}{1978}).

\bibitem[{\citenamefont{Aad et~al.}(2011)}]{Aad:2011hz}
\bibinfo{author}{\bibfnamefont{G.}~\bibnamefont{Aad}} \bibnamefont{et~al.}
  (\bibinfo{collaboration}{ATLAS Collaboration}),
 \bibinfo{journal}{Phys.Lett.} \textbf{\bibinfo{volume}{B703}}, \bibinfo{pages}{428}
  (\bibinfo{year}{2011}), \eprint{1106.4495}.

\bibitem[{\citenamefont{Khachatryan et~al.}(2011)}]{Khachatryan:2011ts}
\bibinfo{author}{\bibfnamefont{V.}~\bibnamefont{Khachatryan}}
  \bibnamefont{et~al.} (\bibinfo{collaboration}{CMS Collaboration}),
  \bibinfo{journal}{JHEP} \textbf{\bibinfo{volume}{1103}}, \bibinfo{pages}{024}
  (\bibinfo{year}{2011}), \eprint{1101.1645}.

\bibitem[{\citenamefont{Eichten and Hill}(1990)}]{Eichten:1989zv}
\bibinfo{author}{\bibfnamefont{E.}~\bibnamefont{Eichten}} \bibnamefont{and}
  \bibinfo{author}{\bibfnamefont{B.~R.} \bibnamefont{Hill}},
  \bibinfo{journal}{Phys.Lett.} \textbf{\bibinfo{volume}{B234}},
  \bibinfo{pages}{511} (\bibinfo{year}{1990}).

\bibitem[{\citenamefont{Grinstein}(1990)}]{Grinstein:1990mj}
\bibinfo{author}{\bibfnamefont{B.}~\bibnamefont{Grinstein}},
  \bibinfo{journal}{Nucl.Phys.} \textbf{\bibinfo{volume}{B339}},
  \bibinfo{pages}{253} (\bibinfo{year}{1990}).

\bibitem[{\citenamefont{Georgi}(1990)}]{Georgi:1990um}
\bibinfo{author}{\bibfnamefont{H.}~\bibnamefont{Georgi}},
  \bibinfo{journal}{Phys.Lett.} \textbf{\bibinfo{volume}{B240}},
  \bibinfo{pages}{447} (\bibinfo{year}{1990}).

\bibitem[{\citenamefont{Falk et~al.}(1990)\citenamefont{Falk, Georgi,
  Grinstein, and Wise}}]{Falk:1990yz}
\bibinfo{author}{\bibfnamefont{A.~F.} \bibnamefont{Falk}},
  \bibinfo{author}{\bibfnamefont{H.}~\bibnamefont{Georgi}},
  \bibinfo{author}{\bibfnamefont{B.}~\bibnamefont{Grinstein}},
  \bibnamefont{and} \bibinfo{author}{\bibfnamefont{M.~B.} \bibnamefont{Wise}},
  \bibinfo{journal}{Nucl.Phys.} \textbf{\bibinfo{volume}{B343}},
  \bibinfo{pages}{1} (\bibinfo{year}{1990}).

\bibitem[{\citenamefont{Chakraverty et~al.}(1996)\citenamefont{Chakraverty, De,
  Dutta-Roy, and Kundu}}]{Chakraverty:1995tb}
\bibinfo{author}{\bibfnamefont{D.}~\bibnamefont{Chakraverty}},
  \bibinfo{author}{\bibfnamefont{T.}~\bibnamefont{De}},
  \bibinfo{author}{\bibfnamefont{B.}~\bibnamefont{Dutta-Roy}},
  \bibnamefont{and} \bibinfo{author}{\bibfnamefont{A.}~\bibnamefont{Kundu}},
  \bibinfo{journal}{Phys.Rev.} \textbf{\bibinfo{volume}{D53}},
  \bibinfo{pages}{5293} (\bibinfo{year}{1996}), \eprint{hep-ph/9504258}.

\bibitem[{\citenamefont{Shifman and Voloshin}(1987)}]{Shifman:1986sm}
\bibinfo{author}{\bibfnamefont{M.~A.} \bibnamefont{Shifman}} \bibnamefont{and}
  \bibinfo{author}{\bibfnamefont{M.}~\bibnamefont{Voloshin}},
  \bibinfo{journal}{Sov.J.Nucl.Phys.} \textbf{\bibinfo{volume}{45}},
  \bibinfo{pages}{292} (\bibinfo{year}{1987}).

\bibitem[{\citenamefont{Shifman and Voloshin}(1988)}]{Shifman:1987rj}
\bibinfo{author}{\bibfnamefont{M.~A.} \bibnamefont{Shifman}} \bibnamefont{and}
  \bibinfo{author}{\bibfnamefont{M.}~\bibnamefont{Voloshin}},
  \bibinfo{journal}{Sov.J.Nucl.Phys.} \textbf{\bibinfo{volume}{47}},
  \bibinfo{pages}{511} (\bibinfo{year}{1988}).

\bibitem[{\citenamefont{Isgur and Wise}(1989)}]{Isgur:1989vq}
\bibinfo{author}{\bibfnamefont{N.}~\bibnamefont{Isgur}} \bibnamefont{and}
  \bibinfo{author}{\bibfnamefont{M.~B.} \bibnamefont{Wise}},
  \bibinfo{journal}{Phys.Lett.} \textbf{\bibinfo{volume}{B232}},
  \bibinfo{pages}{113} (\bibinfo{year}{1989}).

\bibitem[{\citenamefont{Isgur and Wise}(1990)}]{Isgur:1989ed}
\bibinfo{author}{\bibfnamefont{N.}~\bibnamefont{Isgur}} \bibnamefont{and}
  \bibinfo{author}{\bibfnamefont{M.~B.} \bibnamefont{Wise}},
  \bibinfo{journal}{Phys.Lett.} \textbf{\bibinfo{volume}{B237}},
  \bibinfo{pages}{527} (\bibinfo{year}{1990}).

\bibitem[{\citenamefont{Luke and Manohar}(1992)}]{Luke:1992cs}
\bibinfo{author}{\bibfnamefont{M.~E.} \bibnamefont{Luke}} \bibnamefont{and}
  \bibinfo{author}{\bibfnamefont{A.~V.} \bibnamefont{Manohar}},
  \bibinfo{journal}{Phys.Lett.} \textbf{\bibinfo{volume}{B286}},
  \bibinfo{pages}{348} (\bibinfo{year}{1992}), \eprint{hep-ph/9205228}.

\bibitem[{\citenamefont{Chen}(1993)}]{Chen:1993sx}
\bibinfo{author}{\bibfnamefont{Y.-Q.} \bibnamefont{Chen}},
  \bibinfo{journal}{Phys.Lett.} \textbf{\bibinfo{volume}{B317}},
  \bibinfo{pages}{421} (\bibinfo{year}{1993}).

\bibitem[{\citenamefont{Kilian and Ohl}(1994)}]{Kilian:1994mg}
\bibinfo{author}{\bibfnamefont{W.}~\bibnamefont{Kilian}} \bibnamefont{and}
  \bibinfo{author}{\bibfnamefont{T.}~\bibnamefont{Ohl}},
  \bibinfo{journal}{Phys.Rev.} \textbf{\bibinfo{volume}{D50}},
  \bibinfo{pages}{4649} (\bibinfo{year}{1994}), \eprint{hep-ph/9404305}.

\bibitem[{\citenamefont{Sundrum}(1998)}]{Sundrum:1997ut}
\bibinfo{author}{\bibfnamefont{R.}~\bibnamefont{Sundrum}},
  \bibinfo{journal}{Phys.Rev.} \textbf{\bibinfo{volume}{D57}},
  \bibinfo{pages}{331} (\bibinfo{year}{1998}), \eprint{hep-ph/9704256}.

\end{thebibliography}
\end{document}